\documentclass[12pt]{article}
\usepackage{graphicx}

\newcommand\pubnumber{CIPANP2018-Trzaska-374}
\newcommand\pubdate{\today}

\def\Title#1{\begin{center} {\Large #1 } \end{center}}
\def\Author#1{\begin{center}{ \sc #1} \end{center}}
\def\Address#1{\begin{center}{ \it #1} \end{center}}

\newcommand\pubblock{\rightline{\begin{tabular}{l} \pubnumber\\
         \pubdate  \end{tabular}}}
\newenvironment{Abstract}{\begin{quotation}  }{\end{quotation}}
\newenvironment{Presented}{\begin{quotation} \begin{center} 
             PRESENTED AT\end{center}\bigskip 
      \begin{center}\begin{large}}{\end{large}\end{center} \end{quotation}}

\def\beq{\begin{equation}}
\def\eeq#1{\label{#1}\end{equation}}
\def\eeqn{\end{equation}}

\def\beqa{\begin{eqnarray}}
\def\eeqa#1{\label{#1}\end{eqnarray}}
\def\eeqan{\end{eqnarray}}

\let\bar=\overbar

\def\Dslash{\not{\hbox{\kern-4pt $D$}}}
\def\dslash{\not{\hbox{\kern-2pt $\del$}}}

\def\msb{{\bar{\ssstyle M \kern -1pt S}}}

\usepackage[numbers,sort&compress]{natbib}
\usepackage{textcomp}
\usepackage{hyperref}

\begin{document}
\begin{titlepage}
\pubblock

\vfill
\Title{Possibilities for Underground Physics in the~Pyhasalmi mine}
\vfill

\Author{W.H.~Trzaska\textsuperscript{1}, L.~Bezrukov\textsuperscript{2}, T.~Enqvist\textsuperscript{1}, J.~Joutsenvaara\textsuperscript{3}, P.~Kuusiniemi\textsuperscript{1}, K.~Loo\textsuperscript{1}, B.~Lubsandorzhiev\textsuperscript{2}, V.~Sinev\textsuperscript{2}, M.~Slupecki\textsuperscript{1}}

\Address{
{}\textsuperscript{1}Department of~Physics, University of~Jyvaskyla, Finland\\
{}\textsuperscript{2}Institute for Nuclear Research, Russian Academy of~Sciences, Moscow, Russia\\
{}\textsuperscript{3}Kerttu Saalasti Institute, University of~Oulu, Finland}

\vfill
\begin{Abstract}
The Pyhasalmi mine is uniquely suited to host new generation of~large-scale underground experiments. It was chosen both by the~LAGUNA-LBNO and by the~LENA Collaboration as the~preferred site for a~giant neutrino observatory. Regrettably, none of~these projects got funded. The~termination of~the~underground excavations in the~fall of~2019 marks an important milestone. To maintain the~infrastructure in good condition a~new sponsor must be found: either a~large-scale scientific project or a~new commercial operation. The~considered alternatives for the~commercial use of~the~mine include a~pumped-storage hydroelectricity plant and a~high-security underground data-storage centre. Without a~new sponsor the~ongoing experiments, including the~cosmic-ray experiment EMMA and the~study of~C14 content in liquid scintillators, have to be completed within the~next few years. 
\end{Abstract}

\vfill
\begin{Presented}
Thirteenth Conference on the~Intersections\\
of Particle and Nuclear Physics (CIPANP2018)\\
\vfill
Palm Springs, CA, USA,  May 28 - June 3, 2018
\end{Presented}
\vfill
\end{titlepage}
\def\thefootnote{\fnsymbol{footnote}}
\setcounter{footnote}{0}

\section{Introduction}

Earth's atmosphere is exposed to a~steady flux of~energetic particles known as cosmic rays~\cite{Blasi:2013rva_1}. Cosmic-ray interactions with nuclei in the~atmosphere induce intense showers of~secondary particles. Most of~them won't even reach the~ground, but the~high-energy muons created at~the~early stages of~the~shower cascade will penetrate deep under the~surface of~the~Earth. These muons are the~leading source of~background~\cite{Mei:2005gm_2} \mbox{e.g. in experiments} searching for dark matter or aiming to detect neutrinoless double beta decay. To reduce this background, one has to place the~measuring setup deep underground and assure long acquisition times. The~growing interest in low-background measurements generates demand for well-equipped underground laboratory space able to accommodate the~detectors, auxiliary equipment and the~supporting infrastructure.  Such facilities are usually located in unused sections of~a~mine or as an annex to a~road or a~railroad tunnel. The~Pyhasalmi mine in Finland offers an excellent location to host a~variety of~deep-underground experiments. As it was documented during the~LAGUNA-LBNO design study~\cite{Laguna_3}, the~exceptional quality of~the~surrounding rock combined with a~modern mining infrastructure make Pyhasalmi an ideal location for the~construction of~gigantic caverns even at~the~maximum depth of~1400~m below the~ground.

\section{The Pyhasalmi mine}

During the~second half of~2019 the~underground operations in the~Pyhasalmi mine are coming to an end while the~reprocessing activities at~the~surface will still continue for a~few more years. As a~legacy of~nearly 60 years of~operation and especially thanks to the~technological upgrades realized within the~past two decades, the~infrastructure of~the~mine is in excellent condition including modern communication services and safety procedures. The~main level of~the~mine, where all the~major facilities are located, is at~1400~m underground. These facilities, now scheduled for~gradual decommissioning, include four large halls designed for storage, service and maintenance of~mining machinery. There are also control rooms, social areas and a~restaurant. The~1400~m level is accessible from the~ground level by an elevator and by a~\mbox{12-km}~long truck-size decline. The~elevator ride takes about three minutes while a~car ride lasts about half-an-hour. In addition to the~main level, there are additional large caverns at~the depths of~990, 660 and 400~m. 

To promote, maintain, and operate the~underground premises throughout the~transition period and after the~closure of~the~mine, an organization called Callio has been established~\cite{Callio_03}. The~largest of~the~considered commercial alternatives is a~pumped-storage hydroelectricity (PSH) plant~\cite{TeknikkaJaTalous_4}. If realized, it would be the~first PSH in Finland. With the~hydraulic head of~1400~meters and the~storage volume of~162,000~m\textsuperscript{3} it would have the~effective capacity of~1~GWh and utilize one or two 75~MW~turbines. The~other large-scale alternative is an underground data centre. If one or both of~these projects were realized, the~future of~the~mine underground infrastructure would be secured, making a~continuation of~smaller, both scientific and commercial projects possible. 

\begin{figure}[htb]
\centering
\includegraphics[width=6in]{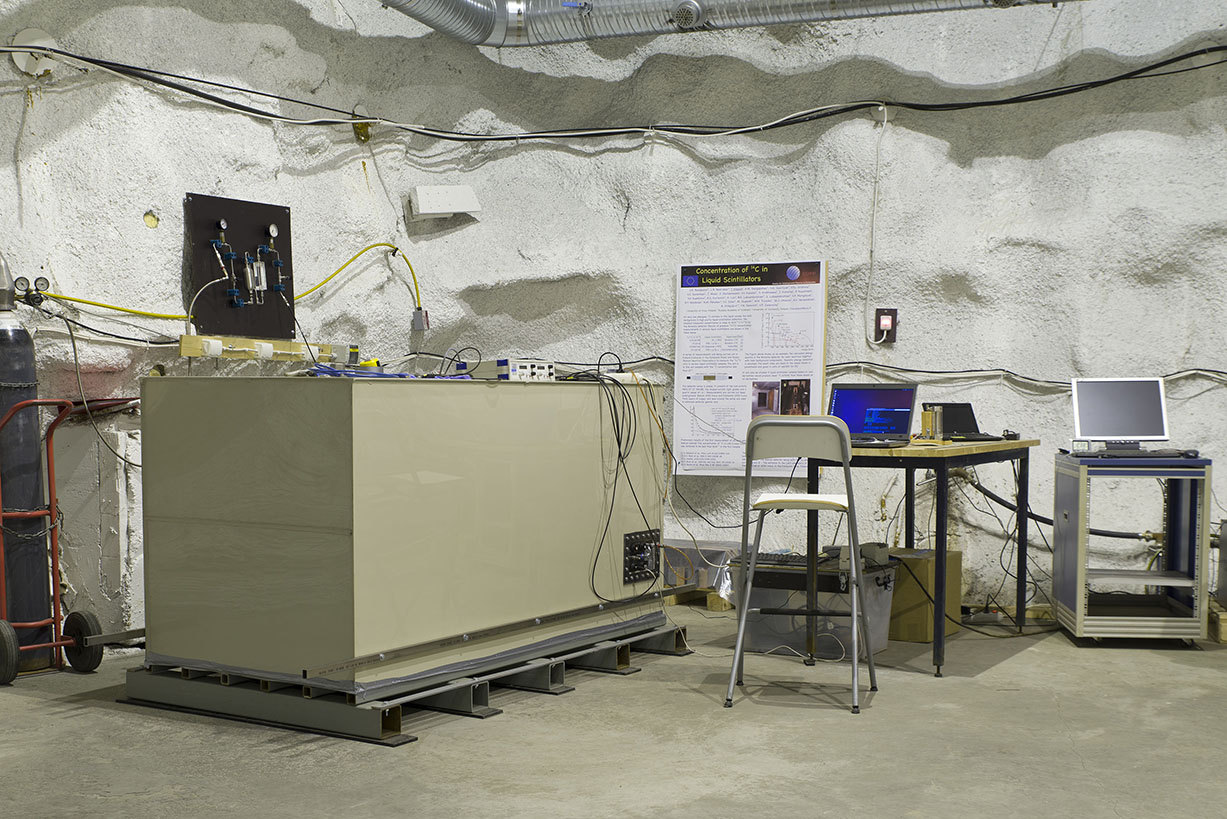}
\caption{A photograph of~the~C14 setup in the~main hall of~Lab2 at~the depth of~1430~m (4100 m.w.e.) in the~Pyhasalmi mine.}
\label{fig-lab-photo}
\end{figure}

\section{Deep-underground laboratory}

The deep-underground laboratory in the~Pyhasalmi, known as Lab2, is located at~the depth of~1430~m corresponding to 4100 m.w.e. The~area of~the~main hall of~Lab2 is about 120 m\textsuperscript{2} and the~maximum height is about 7.5 m. The~area of~the~adjoining entrance hall is about 100~m\textsuperscript{2} and has the~height of~about 5~m, sufficient to accommodate a~delivery truck. Lab2 is situated about 500~m away from the~elevator shaft and is accessible also on foot. The~laboratory space is equipped with the~ventilation, the~optical fibre, and a~1~GB internet connection. Currently the~flux of~fresh air to the~laboratory is maintained at~the~level of~about 10~m\textsuperscript{3}/s. For the~electric power there is a~160~kVA line at~the~entrance hall and a~25~kVA and a~3~kVA (UPS) lines in the~laboratory hall. The~radon level in Lab2 is about 240~Bq/m\textsuperscript{3} and the~temperature is about 26~\textdegree{}C. Figure~\ref{fig-lab-photo} shows the~main hall of~Lab2. The~available space is sufficient to accommodate small- to medium-size experiments and would be well suited e.g. for low-background measurements. 

\subsection{Muon and neutron background in Lab2}

The knowledge of~the~neutron background, originating from radioactive decays and induced by muons, is important for many experiments. The~first muon flux measurements in the~mine date back to 2005. The~measured flux at~the depth of~1390~m (4000~m.w.e.) was $(1.1\pm0.1)\times10^{-4}$ m\textsuperscript{-2}s\textsuperscript{-1}~\cite{Enqvist:2005cy}. The~neutron background measurement in Lab2 was performed in July~2018. The~setup consisted of~\textsuperscript{3}He counters and \textsuperscript{10}\mbox{B-loaded} plastic scintillation detectors. The~results are scheduled to be published soon, but there are already indications that the~flux of~thermal neutrons in the~Pyhasalmi mine is higher than in Gran Sasso. The~current measurements are part of~the~Baltic Sea Underground Innovation Network program (BSUIN)~\cite{Bsuin_5}.

\section{Ongoing physics experiments in the~mine}

Currently there are two physics experiments taking data in the~mine: the~cosmic-ray experiment EMMA at~the depth of~75~m and the~C14 experiment probing the~concentration of~\textsuperscript{14}C in liquid scintillators in the~main hall of~Lab2.

\subsection{EMMA}

\begin{figure}[htb]
\centering
\includegraphics[width=5in]{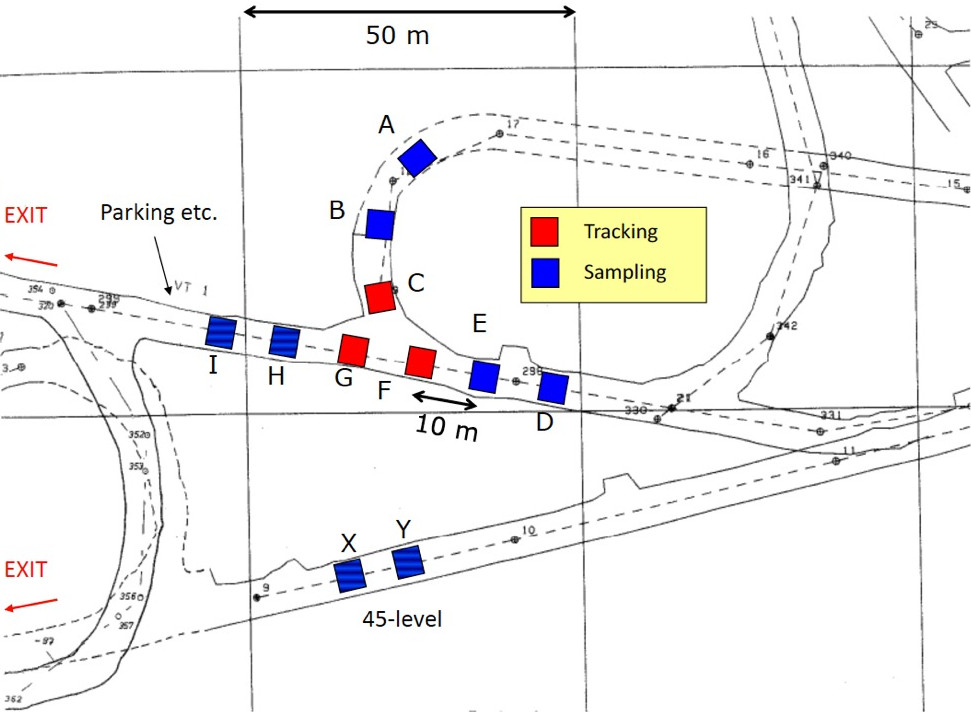}
\caption{Layout of~detector stations of~the~EMMA experiment. The~three central stations provide high-resolution tracking to determine the~direction of~the~muon shower around the~core. The~remaining stations have adequate resolution to sample the~density at~the~edges of~the~distribution.}
\label{fig-EMMA}
\end{figure}

EMMA (Experiment with Multi-Muon Array)~\cite{Kuusiniemi:2018vbz_6} is a~dedicated underground cosmic-ray experiment studying the~mass composition at~the~knee region. The~setup consists of~11 nearly completed detector stations situated at~the~shallow depth of~75~m. This overburden of~about 210~m.w.e. provides about 45~GeV cut-off energy for atmospheric muons. The~three central stations are capable of~performing muon tracking with the~angular resolution of~up to 1~degree and with density resolution of~up to 60~muons~per~m\textsuperscript{2}. The~remaining stations (Fig.~\ref{fig-EMMA}) are intended to sample the~lateral density distribution of~muons over the~fiducial area of~about~300~m\textsuperscript{2}. 

Our simulations indicate that the~lateral density distributions of~muons with energy over 45~GeV are model-independent and contain the~information both on the~mass and on the~energy of~the~primary cosmic ray~\cite{Sarkamo:2014_7}. The~energy can be deduced from the~muon density at~the~core while the~mass -- from the~slope of~the~lateral distribution.

If completed and operated for three years in the~full configuration, EMMA would make a~significant contribution towards solving the~long-standing puzzle of~the~presence of~the~knee in the~energy spectrum of~cosmic rays. This, in turn, may reveal further information on the~cosmic-ray sources and acceleration mechanisms. The~current understanding is that the~acceleration can take place up to the~knee energies in supernova shock fronts that could propagate thousands of~years after the~explosion. However, there should be also other mechanisms since the~supernova shock front mechanism does not produce energies above the~knee. Addressing these questions is relevant and important as they are among the~major topics in the~present-day astrophysics.

The second important task for EMMA is to verify the~alleged muon excess~\cite{Bellido:2018toz_8} in the~extensive air showers (EAS). This problem is extremely important because, if confirmed, it would force substantial revision of~the~existing particle interaction models with serious repercussions in multiple fields of~science relaying on these models.

\section{C14 experiment}

The beta decay of~the~long-lived radioactive \textsuperscript{14}C is the~main source of~background for low-energy ($E\simeq300$~keV) neutrino measurements using high-purity liquid scintillation detectors~\cite{Wurm:2011zn_9}. The~lowest \textsuperscript{14}C concentration has been reported by the~Borexino Collaboration for Pseudocumene (PC) amounting to $\sim2\times10^{-18}$~\cite{Alimonti:1998rc_10}. There are three other published results for the~concentration (for PXE and PC+Dodecane) with the~highest being $(12.6\pm0.4)\times10^{-18}$~\cite{Back:2008zz_11}\cite{Keefer:2011kf_12}\cite{Buck:2012zz_13}. Currently the~preferred solvent for the~new generation of~large neutrino detectors is LAB (Linear alkylbenzene). LAB, just like the~other petrochemical products, is synthesized from the~crude gas or oil extracted from deep geological deposits where the~expected remanences of~the~cosmogenic \textsuperscript{14}C are very low. It should therefore be possible~\cite{Bonvicini:2003zw_14}, by a~careful selection of~the~gas field that is free of~recent contaminants, to reach low concentration of~the~radiocarbon. The~concentration of~\textsuperscript{14}C in LAB has not yet been measured with sufficient sensitivity. Lab2 in the~Pyhasalmi mine is one of~the~sites where a~dedicated setup for such measurements is being constructed (Fig.~\ref{fig-lab-photo}). It is intended to make systematic analysis of~the~samples of~different origin and composition with the~aim of~finding concentrations smaller than $10^{−18}$ for the~use, for instance, by the~SNO+~\cite{Chen:2008un_15} and the~JUNO Collaboration~\cite{An:2015jdp_16}. Such low concentrations are currently below the~sensitivity of~the~Atomic Mass Spectrometry~\cite{ams_015}.

\section{Future plans and possibilities}
\subsection{Giant liquid-based neutrino detectors}

Considering the~modern infrastructure and safety, the~quality of~the~surrounding rock, the~small footprint of~the~ore deposit, the~low maintenance costs, and the~depth of~the~main level (4100~m.w.e.), the~Pyhasalmi mine has ideal conditions to host underground experiments of~the~next generation~\cite{Trzaska:2011zz_17}. In fact, Pyhasalmi was already selected as the~prime site for the~far detector of~the~LAGUNA-LBNO project. The~plan was to produce neutrino beam at~CERN and send it over the~distance of~2288 km to Finland~\cite{Galymov:2016nwz_18}. However, following new strategy and cooperation agreements, CERN has ceded to Fermilab all accelerator-based neutrino physics research and LAGUNA was replaced by the~DUNE experiment~\cite{Dune_19}. Nevertheless, the~feasibility for the~construction of~giant caverns, capable of~containing 50~kton detectors together with the~needed equipment in a~single cave has been confirmed~\cite{Trzaska:2012jt_20}.

The second of~the~LAGUNA detectors that has chosen the~Pyhasalmi mine as its preferred location is LENA (Low Energy Neutrino Astronomy)~\cite{Wurm:2011zn_9} -- a~50~kton liquid scintillator, multi-purpose neutrino observatory. Unfortunately, the~LENA Collaboration failed to obtain adequate support from the~funding agencies to realize the project. Instead, the~majority of~the~neutrino scientists involved with the~liquid scintillator technology have joined the~reactor neutrino experiment -- JUNO~\cite{An:2015jdp_16}. Nevertheless, the~bulk of~the~astroparticle goals of~LENA cannot be covered by JUNO as it is located at~a~relatively shallow depth of~600~meters and at~the~distance of~53~km from 10 high-power nuclear reactors. It is therefore conceivable that, in a~few years and benefiting from the~technological developments of~JUNO, LENA or a~similar project will be reconsidered.

\subsection{Acoustic detection of~neutrinos in the~rock}

\begin{figure}[htb]
\centering
\includegraphics[width=4in]{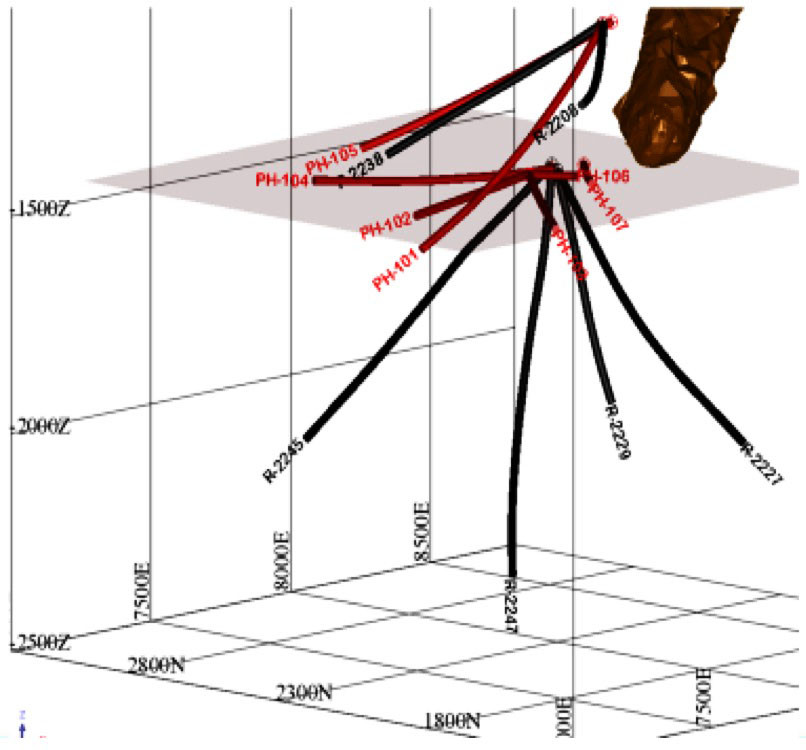}
\caption{A network of~3.5~km of~boreholes covering the~volume of~about~1~km\textsuperscript{3} extending from the~depth of~around 1300~m down to~2500~m. It has been proposed to use the~boreholes for the~deployment of~microphones for acoustic detection of~\mbox{high-energy} neutrinos.}
\label{fig-boreholes}
\end{figure}

One of~the~requirements of~the~extended site investigation for LAGUNA-LBNO was to drill a~network of~boreholes reaching far out and deep down from the~vicinity of~the~main level into the~surrounding area. These boreholes have very well documented geological profile and are now available for scientific research. As illustrated in the~Fig.~\ref{fig-boreholes}, the~explored area covers the~volume of~about~1~km\textsuperscript{3} reaching from the~depth of~around 1300~m down to~2500~m. The~total length of~the~new boreholes is~3.5~km. It has been proposed~\cite{Trzaska:2018_21} to deposit strings of~microphones into the~boreholes in a~similar fashion it has been done or is going to be done by the~ANTARES/AMADEUS~\cite{Aguilar:2010ac_22} and the~KM3NeT~\cite{Adrian-Martinez:2016fdl_23} collaborations for the~purpose of~acoustic detection of~particle showers following interactions of~ultra-high energy neutrinos. This type of~measurement in the~rock have never been tried or even proposed before. Nevertheless, since the~density of~rock is three-times larger and the~speed of~sound is four-times larger, the~amplitude of~the~generated bipolar pressure pulse in rock should be by an order of~magnitude larger than in water. In addition, a~higher density of~rock also guarantees higher interaction rate for neutrinos while a~longer attenuation length in rock reduces signal dissipation. 

\section{Conclusions}

The Pyhasalmi mine is uniquely suited to host new generation of~large-scale underground experiments. It was chosen both by the~LAGUNA-LBNO and by the~LENA Collaboration as the~preferred site for a~giant neutrino observatory. Regrettably, none of~these projects got funded. The~termination of~the~underground excavations in the~fall of~2019 marks an important milestone. To maintain the~underground infrastructure a~new sponsor must be found: either a~large-scale scientific project or new commercial operation. The~currently considered alternatives for the~commercial use of~the~mine include a~pumped-storage hydroelectricity plant and a~high-security underground data-storage centre. Without a~new sponsor the~ongoing experiments, including the~cosmic-ray experiment EMMA and the~study of~\textsuperscript{14}C content in liquid scintillators, have to be completed within the~next few years.


\bibliography{eprint}


\end{document}